\documentclass{article}%
\usepackage{amsfonts}
\usepackage{amsmath}
\usepackage{amssymb}
\usepackage{charter}
\usepackage{graphicx}%
\providecommand{\U}[1]{\protect \rule{.1in}{.1in}}

\providecommand{\U}[1]{\protect \rule{.1in}{.1in}}

\begin{document}

\title{Nonclassicality of photon-added squeezed vacuum and its decoherence in thermal
environment{\small \thanks{Project supported by the National Natural Science
Foundation of China (Grant Nos 10775097 and 10874174) and the Research
Foundation of the Education Department of Jiangxi Province.}}}
\author{{\small Li-yun Hu}$^{1,2}${\small \thanks{Corresponding author. \emph{E-mail
address}: hlyun2008@126.com (L-Y Hu).} and Hong-yi Fan}$^{2}$\\$^{1}${\small College of Physics and Communication Electronics, Jiangxi Normal
University, Nanchang 330022, China}\\$^{2}${\small Department of Physics, Shanghai Jiao Tong University, Shanghai,
200030, China}}
\maketitle

\begin{abstract}
{\small We study the nonclassicality of photon-added squeezed vacuum (PASV)
and its decoherence in thermal environment in terms of the sub-Poissonian
statistics and the negativity of Wigner function (WF). By converting the PASV
to a squeezed Hermite polynomial excitation state, we derive a compact
expression for the normalization factor of }$m$-{\small PASV, which is an }%
$m${\small -order Legendre polynomial of squeezing parameter} $r${\small . We
also derive the explicit expression of WF of }$m$-{\small PASV and find the
negative region of WF in phase space. We show that there is an upper bound
value of }$r$ {\small for this state to exhibit sub-Poissonian statistics
increasing as }$m$ {\small increases. Then we} {\small derive the explicit
analytical expression of time evolution of WF of }$m${\small -PASV in the
thermal channel and discuss the loss of nonclassicality using the negativity
of WF. The threshold value of decay time is presented for the single PASV.}

{\small PACS number(s): 03.67.-a, 03.65. Ud, 42. 50.Dv}

\end{abstract}

\section{Introduction}

Nonclassical states play an important role in quantum computation and quantum
imformation procession \cite{1}. The nonclassicality of quantum states can be
well-described by some nonclassical properties, such as sub-Poissonian photon
ststistics \cite{2}, squeezing in one of the quadratures of the field
\cite{3}, and negativity of Wigner function (WF) \cite{4}. Especially, the
partial negativity of WF implies the highly nonclassical properties of quantum
states and is often used to describe the decoherence of quantum states. Many
experimental schemes have been proposed to generate nonclassical states of
optical field. Among them, subtracting photons form and/or adding photons to
quantum states have been paid much attention because these fields exhibit an
abundant of nonclassical properties and may give access to a complete
engineering of quantum states and to fundamental quantum phenomena
\cite{5,6,7,8,9,10,11,12}. For example, a single photon-subtraction squeezed
vacuum (PSSV) has been experimentally prepared with a pulsed and continuous
wave squeezed vacuum \cite{10,11,12}. It is very similar to quantum
superpositions of coherent states with small amplitudes \cite{13,14}. As
another example, a single-photon addition was experimentally performed by
Zavatta \textit{et al} \cite{6}, which unveils the nonclassical features
associated with the excitation of a classical coherent field by a single light
quantum. For the single PSSV, its nonclassical properties and decoherence was
investigated theoretically in two different decoherent channels (amplitude
decay and phase damping) by Biswas and Agarwal \cite{15}. They indicated that
the WF losses its non-Gaussian nature and becomes Gaussian at long times in
amplitude decay case.

On the other hand, the combination of the photon addition and subtraction has
been successfully demonstrated in Ref.\cite{8,8a}. In Ref.\cite{8}, photon
addition and subtraction experimentally have been employed to probe quantum
commutation rules by Parigi \textit{et al}. In fact, they have implemented
simple alternated sequences of photon creation (addition) and annihilation
(subtraction) on a thermal field and observed the noncommutativity of the
creation and annihilation operators. It is interesting to notice that
subtracting or adding one photon from/to pure squeezed vacuums can generate
the same output state, i.e., squeezed single-photon state \cite{16}. However,
the resulting states obtained by successive photon subtractions or additions
are different from each other. For instance, successive two-photon additions
[$a^{\dag2}$] and successive two-photon subtractions [$a^{2}$] will result in
the same state produced by using subtraction-addition ($a^{\dag}a$) and
addition-subtraction ($aa^{\dag}$), (also see section 2 below) respectively.
As far as we know, the nonclassicality and decoherence of arbitrary number
photon-added squeezed vacuum states (PASV) in a dissipative channel has not
been derived analytically in the literature before.

In this paper, we shall investigate the nonclassical properties and
decoherence of single-mode PASV which is optically produced single-mode
non-Gaussian states. This work is arranged as follows. In section 2, we
introduce the single-mode PASV and discuss its nonclassicality in terms of
sub-Poissonian statistics and the negativity of its Wigner function (WF). By
converting the PASV to a squeezed Hermite polynomial excitation state, we
derive a compact expression for the normalization factor of PASV, which is an
$m$-order Legendre polynomial of the squeezing parameter $r$, where $m$ is the
number of added photons; and then derive the explicit analytical expression of
WF for any photon-added number $m$ and find the negative region of WF in phase
space. We also show that there is an upper bound value of $r$ for this state
to exhibit sub-Poissonian statistics which increases as $m$ increases. In
section 3, we derive the explicit analytical expression of time evolution of
WF of the arbitrary PASV in the thermal channel and discuss the loss of
nonclassicality in reference of the negativity of WF. The threshold value of
the decay time corresponding to the transition of the WF from partial negative
to completely positive definite is obtained at the center of the phase space,
which is independent of the squeezing parameter. We show that the WF for
single PASV has always negative value for all parameters $r$ if the decay time
$\kappa t<\frac{1}{2}\ln \frac{2\bar{n}+2}{2\bar{n}+1}\ $(see Eq.(\ref{4.16})
below), where $\bar{n}$ denotes the average thermal photon number in the
environment\textbf{\ }with dissipative coefficient\textbf{\ }$\kappa$.
Conclusions are involved in the last section.

\section{Single-mode Photon added squeezed vacuum state}

Various photon states have been generated by the micromaser and WFs of some
cavity fields can be measured by a scheme based on interaction between cavity
fields and atoms \cite{16a,16b}. As described in Ref. \cite{16c} when an
excited atom passes through a cavity field which is in a squeezed vacuum state
then their interaction may produce photon addition (excitation) on the
squeezed vacuum state---the excited squeezed vacuum state.

Therectically, the single-mode PASV can be obtained by repeatedly operating
the photon creation operator $a^{\dag}$\ on a squeezed vacuum state $S\left(
r\right)  \left \vert 0\right \rangle $, i.e.,
\begin{equation}
\left \vert r,m\right \rangle \equiv N_{r,m}a^{\dag m}S\left(  r\right)
\left \vert 0\right \rangle , \label{2.1}%
\end{equation}
where $\left \vert 0\right \rangle $ is single mode vacuum, and $N_{r,m}\ $is
the normalization constant to be determined, $a^{\dag}$ is the Bose creation
operator, and $S\left(  r\right)  $ is the single-mode squeezing operator
$S\left(  \lambda \right)  =\exp[\frac{1}{2}\left(  ra^{\dagger2}%
-ra^{2}\right)  ]$ \cite{17,18} with $r$ being the squeezing parameter.

\subsection{Single-mode PASV as the squeezed Hermite polynomial excitation
state}

Under the transform of $S\left(  \lambda \right)  $ we see $S^{\dag}\left(
r\right)  a^{\dagger}S\left(  r\right)  =a^{\dagger}\cosh r+a\sinh r,$ thus we
can reform Eq.(\ref{2.1}) as%
\begin{align}
\left \vert r,m\right \rangle  &  =N_{r,m}S\left(  r\right)  S^{\dag}\left(
r\right)  a^{\dag m}S\left(  r\right)  \left \vert 0\right \rangle \nonumber \\
&  =N_{r,m}S\left(  r\right)  \left(  a^{\dag}\cosh r+a\sinh r\right)
^{m}\left \vert 0\right \rangle . \label{2.2}%
\end{align}
On the other hand, using the operator identity \cite{19}%
\begin{equation}
\left(  a\mu+\nu a^{\dagger}\right)  ^{m}=\left(  -i\sqrt{\frac{\mu \nu}{2}%
}\right)  ^{m}\colon H_{m}\left(  i\sqrt{\frac{\mu}{2\nu}}a+i\sqrt{\frac{\nu
}{2\mu}}a^{\dag}\right)  \colon, \label{2.3}%
\end{equation}
where $H_{m}\left(  x\right)  $ is $m$-order single variable Hermite
polynomial whose definition is
\[
H_{m}\left(  x\right)  =\sum_{l=0}^{[m/2]}\frac{\left(  -1\right)  ^{l}%
m!}{l!(m-2l)!}\left(  2x\right)  ^{m-2l},
\]
we have%
\begin{align}
\left \vert r,m\right \rangle  &  =\frac{\left(  -i\right)  ^{m}}{2^{m}}%
N_{r,m}\sinh^{m/2}2rS\left(  r\right)  \colon H_{m}\left(  i\sqrt{\frac{\tanh
r}{2}}a+i\sqrt{\frac{\coth r}{2}}a^{\dag}\right)  \colon \left \vert
0\right \rangle \nonumber \\
&  =\frac{\left(  -i\right)  ^{m}}{2^{m}}N_{r,m}\sinh^{m/2}2rS\left(
r\right)  H_{m}\left(  i\sqrt{\frac{\coth r}{2}}a^{\dag}\right)  \left \vert
0\right \rangle , \label{2.4}%
\end{align}
which indicates that the PASV is equivalent to a squeezed Hermite-excited
vacuum state.

Further using the generating function of $H_{m}\left(  x\right)  $,
\begin{equation}
H_{m}\left(  x\right)  =\left.  \frac{\partial^{m}}{\partial t^{m}}\exp \left(
2xt-t^{2}\right)  \right \vert _{t=0}, \label{2.5}%
\end{equation}
and Eq.(\ref{2.4}), the normalization factor $N_{r,m}$ can be derived by
\begin{align}
1  &  =\frac{N_{r,m}^{2}}{2^{2m}}\sinh^{m}2r\left \langle 0\right \vert
H_{m}\left(  -i\sqrt{\frac{\coth r}{2}}a\right)  H_{m}\left(  i\sqrt
{\frac{\coth r}{2}}a^{\dag}\right)  \left \vert 0\right \rangle \nonumber \\
&  =\frac{N_{r,m}^{2}}{2^{2m}}\sinh^{m}2r\frac{\partial^{2m}}{\partial \tau
^{m}\partial t^{m}}e^{-t^{2}-\tau^{2}}\left.  \left \langle 0\right \vert
e^{-i\sqrt{2\coth r}a\tau}e^{i\sqrt{2\coth r}a^{\dag}t}\left \vert
0\right \rangle \right \vert _{\tau=t=0}\nonumber \\
&  =\frac{N_{r,m}^{2}}{2^{2m}}\sinh^{m}2r\frac{\partial^{2m}}{\partial \tau
^{m}\partial t^{m}}\left.  \exp \left(  -t^{2}-\tau^{2}+2\tau t\coth r\right)
\right \vert _{\tau=t=0}, \label{2.6}%
\end{align}
where in the last step we have used the Baker-Hausdorff formula $e^{\mu
a}e^{\nu a^{\dagger}}=e^{\nu a^{\dagger}}e^{\mu a}e^{\mu \nu}$. Using the newly
found generating function of Legendre polynomial \cite{20,21} (see Appendix
A),%
\begin{equation}
\frac{\partial^{2m}}{\partial t^{m}\partial \tau^{m}}\left.  \exp \left(
-t^{2}-\tau^{2}+\frac{2x\tau t}{\sqrt{x^{2}-1}}\right)  \right \vert
_{t,\tau=0}=\frac{2^{m}m!}{\left(  x^{2}-1\right)  ^{m/2}}P_{m}\left(
x\right)  , \label{2.7}%
\end{equation}
we can derive the compact expression for $N_{r,m}$, i.e.,
\begin{equation}
N_{r,m}^{-2}=m!\cosh^{m}rP_{m}\left(  \cosh r\right)  , \label{2.8}%
\end{equation}
which is just the result in Ref.\cite{16c} derived by the mathematical
induction method. In particular, when $m=1,$ ($H_{1}\left(  x\right)  =2x$)
i.e., the single-photon added case, we see that $\left \vert r,1\right \rangle
=S\left(  r\right)  a^{\dag}\left \vert 0\right \rangle $ is just a squeezed
single-photon state; on the other hand, for the single-photon subtracted case
\cite{16,23}, the state is $aS\left(  r\right)  \left \vert 0\right \rangle
=S\left(  r\right)  S^{\dagger}\left(  r\right)  aS\left(  r\right)
\left \vert 0\right \rangle =S\left(  r\right)  \left(  a\cosh r+a^{\dag}\sinh
r\right)  \left \vert 0\right \rangle \rightarrow S\left(  r\right)  a^{\dag
}\left \vert 0\right \rangle ,$ which indicates that adding a single-photon to
the squeezed state has the same impact as annihilating a photon from the
squeezed state. While for $m\geqslant2,$ the case is not true (also see
Fig.1). For example, successive two-photon additions [$a^{\dag2}$] and
successive two-photon subtractions [$a^{2}$] will result in the same state
produced by using subtraction-addition ($a^{\dag}a$) and addition-subtraction
($aa^{\dag}$), respectively, i.e., $a^{\dag2}S\left(  r\right)  \left \vert
0\right \rangle \rightarrow a^{\dag}aS\left(  r\right)  \left \vert
0\right \rangle ,a^{2}S\left(  r\right)  \left \vert 0\right \rangle \rightarrow
aa^{\dag}S\left(  r\right)  \left \vert 0\right \rangle .$ In Ref.\cite{24}, two
PSSV is used to generate the squeezed superposition of coherent states with
high fidelities and large amplitudes.

Combining Eqs.(\ref{2.1}) and (\ref{2.8}) we can conveniently calculate the
average photon number $a^{\dag}a$ in PASV,
\begin{align}
\left \langle r,m\right \vert a^{\dag}a\left \vert r,m\right \rangle  &
=\left \langle r,m\right \vert aa^{\dag}\left \vert r,m\right \rangle
-1\nonumber \\
&  =\frac{N_{r,m}^{2}}{N_{r,m+1}^{2}}-1\nonumber \\
&  =\left(  m+1\right)  \zeta \frac{P_{m+1}\left(  \zeta \right)  }{P_{m}\left(
\zeta \right)  }-1,\label{2.9}%
\end{align}
where we denote $\zeta=\cosh r$ for simplicity, and%
\begin{align}
\left \langle r,m\right \vert a^{\dag2}a^{2}\left \vert r,m\right \rangle  &
=\left \langle r,m\right \vert \left(  a^{2}a^{\dag2}-4aa^{\dag}+2\right)
\left \vert r,m\right \rangle \nonumber \\
&  =\frac{N_{r,m}^{2}}{N_{r,m+2}^{2}}-4\frac{N_{r,m}^{2}}{N_{r,m+1}^{2}%
}+2\nonumber \\
&  =\left(  m+1\right)  \zeta \left \{  \left(  m+2\right)  \zeta \frac
{P_{m+2}\left(  \zeta \right)  }{P_{m}\left(  \zeta \right)  }-4\frac
{P_{m+1}\left(  \zeta \right)  }{P_{m}\left(  \zeta \right)  }\right \}
+2,\label{2.10}%
\end{align}
thus the Mandel's $\mathcal{Q}$-parameter can be obtained by substituting
Eqs.(\ref{2.9}) and (\ref{2.10}) into $\mathcal{Q}\equiv \frac{\left \langle
a^{\dagger2}a^{2}\right \rangle }{\left \langle a^{\dag}a\right \rangle
}-\left \langle a^{\dag}a\right \rangle $. In particular, for
single-photon-added case $m=1,$ Eqs.(\ref{2.9}) and (\ref{2.10}) reduce to
\begin{align}
\left \langle r,1\right \vert a^{\dag}a\left \vert r,1\right \rangle  &
=3\cosh^{2}r-2,\label{2.11}\\
\left \langle r,1\right \vert a^{\dag2}a^{2}\left \vert r,1\right \rangle  &
=\frac{3\left(  3+2\tanh^{2}r\right)  }{\left(  \coth r-\tanh r\right)  ^{2}%
},\label{2.12}%
\end{align}
thus the $\mathcal{Q}$-parameter with $m=1$ is given by%
\begin{equation}
\mathcal{Q=}\frac{3\sinh^{2}2r}{3\cosh2r-1}-1.\label{2.13}%
\end{equation}
From Eq.(\ref{2.13}) we find that $\mathcal{Q}$ becomes negative for $m=1$
which is satisfied for the squeezing parameter $r\lesssim0.46$ similar to the
result of $\mathcal{Q}$ for single-photon subtracted case \cite{15}. In order
to see clearly the variation of $\mathcal{Q}$-parameter with $r$, we show the
plots of $\mathcal{Q}$-parameter in Fig.1, from which one can clearly see that
$\mathcal{Q}$-parameter\ becomes negetive ($m\neq0)$ when $r$ is less than a
certain threshold value which increases as $m$ increases; while for $m=0,$
$\mathcal{Q}$ is always positive. This implies that the nonclassicality is
enhanced by adding photon to squeezed state. We should emphasize that the WF
has negative region for all $r,$ and thus the PASV is nonclassical. In our
following work, we pay attention to the (ideal) PASV in a thermal channel.

\begin{figure}[ptb]
\label{Fig0} \centering
\includegraphics[width=10cm]{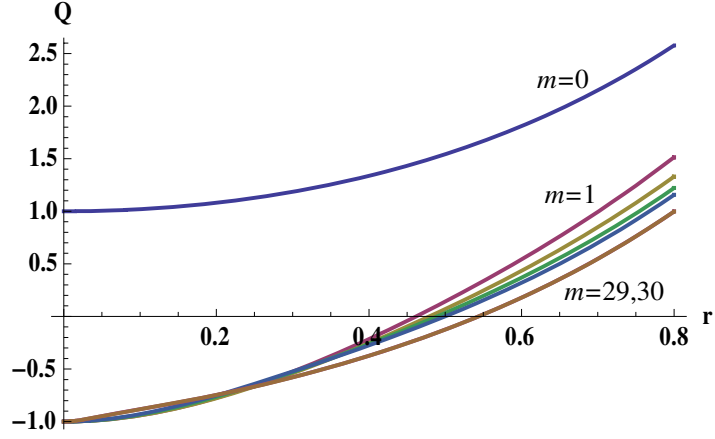}\caption{{\protect \small (Color online)
The }$Q${\protect \small -parameter as the function of squeezing parameter }%
$r${\protect \small \ for different }$m=0,1,2,3,4,29,30.$}%
\end{figure}

\subsection{Wigner function of PASV}

In order to discuss the decoherence properties of PASV in thermal environment,
in this subsection we shall derive the analytical expression of Wigner
function for PASVS. For single-mode case, the Wigner operator is defined as
\cite{25}
\begin{equation}
\Delta \left(  q,p\right)  =\frac{1}{2\pi}\int_{-\infty}^{\infty}du\left \vert
q-\frac{u}{2}\right \rangle \left \langle q+\frac{u}{2}\right \vert
e^{-ipu},\label{3.1}%
\end{equation}
where $\left \vert q\right \rangle $ is the coordinate representation,
$Q\left \vert q\right \rangle =q\left \vert q\right \rangle $. Thus the Wigner
function of PASVS $\left \vert r,m\right \rangle $ can be calculated by
$W\left(  q,p\right)  =\left \langle r,m\right \vert \Delta \left(  q,p\right)
\left \vert r,m\right \rangle .$ Using Eq.(\ref{2.4}), we can see
\begin{equation}
W\left(  q,p\right)  =\frac{N_{r,m}^{2}}{2^{2m}}\sinh^{m}2r\left \langle
0\right \vert H_{m}\left(  -i\sqrt{\frac{\coth r}{2}}a\right)  S^{\dag}\left(
r\right)  \Delta \left(  q,p\right)  S\left(  r\right)  H_{m}\left(
i\sqrt{\frac{\coth r}{2}}a^{\dag}\right)  \left \vert 0\right \rangle
.\label{3.2}%
\end{equation}
On the other hand, noticing that the single-mode squeezing operator $S^{\dag
}\left(  r\right)  $ has its natural expression in coordinate representation
\cite{26}, i.e., $S^{\dag}\left(  r\right)  =e^{r/2}\int_{-\infty}^{\infty
}dq\left \vert qe^{r}\right \rangle \left \langle q\right \vert ,$ leading to
$S^{\dag}\left(  r\right)  \left \vert q\right \rangle =e^{-r/2}\left \vert
qe^{-r}\right \rangle ,$ it then follows that
\begin{align}
&  S^{\dag}\left(  r\right)  \Delta \left(  q,p\right)  S\left(  r\right)
\nonumber \\
&  =\frac{1}{2\pi}\int_{-\infty}^{\infty}d\left(  ue^{-r}\right)  \left \vert
qe^{-r}-\frac{ue^{-r}}{2}\right \rangle \left \langle qe^{-r}+\frac{ue^{-r}}%
{2}\right \vert e^{-ipu}\nonumber \\
&  =\Delta \left(  qe^{-r},pe^{r}\right)  .\label{3.3}%
\end{align}
Then substituting Eq.(\ref{3.3}) into Eq.(\ref{3.2}) and using Eq.(\ref{2.5})
as well as the coherent state representation of Wigner operator,
\begin{equation}
\Delta \left(  q,p\right)  \rightarrow \Delta \left(  \alpha,\alpha^{\ast
}\right)  =\frac{e^{2\left \vert \alpha \right \vert ^{2}}}{\pi}\int \frac{d^{2}%
z}{\pi}\left \vert z\right \rangle \left \langle -z\right \vert e^{-2\left(
z\alpha^{\ast}-z^{\ast}\alpha \right)  },\label{3.4}%
\end{equation}
where $\alpha=(q+ip)/\sqrt{2}$ and $\left \vert z\right \rangle =\exp
(za^{\dagger}-z^{\ast}a)\left \vert 0\right \rangle $ is the coherent state
\cite{27,28}, we can put Eq.(\ref{3.2}) into the following form,%
\begin{align}
W\left(  q,p\right)   &  =\frac{N_{r,m}^{2}}{2^{2m}}\sinh^{m}2r\left \langle
0\right \vert H_{m}\left(  -i\sqrt{\frac{\coth r}{2}}a\right)  \Delta \left(
qe^{-r},pe^{r}\right)  H_{m}\left(  i\sqrt{\frac{\coth r}{2}}a^{\dag}\right)
\left \vert 0\right \rangle \nonumber \\
&  =\frac{N_{r,m}^{2}}{2^{2m}}\sinh^{m}2r\frac{\partial^{2m}}{\partial
t^{m}\partial \tau^{m}}e^{-t^{2}-\tau^{2}}\left.  \left \langle 0\right \vert
e^{-i\sqrt{2\coth r}at}\Delta \left(  qe^{-r},pe^{r}\right)  e^{i\sqrt{2\coth
r}a^{\dag}\tau}\left \vert 0\right \rangle \right \vert _{\tau=t=0}\nonumber \\
&  =\frac{N_{r,m}^{2}e^{2\left \vert \beta \right \vert ^{2}}}{2^{2m}\pi}%
\sinh^{m}2r\frac{\partial^{2m}}{\partial t^{m}\partial \tau^{m}}e^{-t^{2}%
-\tau^{2}}\left.  \int \frac{d^{2}z}{\pi}e^{-\left \vert z\right \vert
^{2}-\left(  i\sqrt{2\coth r}t+2\beta^{\ast}\right)  z+\left(  2\beta
-i\sqrt{2\coth r}\tau \right)  z^{\ast}}\right \vert _{\tau=t=0}\nonumber \\
&  =\frac{N_{r,m}^{2}e^{-2\left \vert \beta \right \vert ^{2}}}{2^{2m}\pi}%
\sinh^{m}2r\left.  \frac{\partial^{2m}}{\partial t^{m}\partial \tau^{m}%
}e^{-t^{2}+2\bar{\beta}t-\tau^{2}+2\allowbreak \bar{\beta}^{\ast}\tau
-2t\tau \coth r}\right \vert _{\tau=t=0},\label{3.5}%
\end{align}
where
\begin{equation}
\bar{\beta}=-i\beta \sqrt{2\coth r},\beta=(qe^{-r}+ipe^{r})/\sqrt{2}%
=\alpha \cosh r-\allowbreak \alpha^{\ast}\sinh r,\label{3.5b}%
\end{equation}
and in the last step, we have used the integration formula
\begin{equation}
\int \frac{d^{2}z}{\pi}e^{\zeta \left \vert z\right \vert ^{2}+\xi z+\eta z^{\ast
}}=-\frac{1}{\zeta}e^{-\frac{\xi \eta}{\zeta}},\text{Re}\left(  \zeta \right)
<0.\label{3.6}%
\end{equation}
In order to further simplify Eq.(\ref{3.5}), expanding the exponential item
$e^{-2t\tau \coth r}$ as sum series and using Eq.(\ref{2.5}) we have%
\begin{align}
W\left(  q,p\right)   &  =\frac{N_{m}^{2}e^{-2\left \vert \beta \right \vert
^{2}}}{2^{2m}\pi}\sinh^{m}2r\sum_{l=0}^{\infty}\frac{\left(  -2\coth r\right)
^{l}}{2^{2l}l!}\frac{\partial^{2l}}{\partial \left(  \bar{\beta}\right)
^{l}\partial \left(  \bar{\beta}^{\ast}\right)  ^{l}}\left.  \frac
{\partial^{2m}}{\partial t^{m}\partial \tau^{m}}e^{-t^{2}+2\bar{\beta}%
t-\tau^{2}+2\bar{\beta}^{\ast}\tau}\right \vert _{\tau=t=0}\nonumber \\
&  =\frac{N_{m}^{2}e^{-2\left \vert \beta \right \vert ^{2}}}{2^{2m}\pi}\sinh
^{m}2r\sum_{l=0}^{\infty}\frac{\left(  -2\coth r\right)  ^{l}}{2^{2l}l!}%
\frac{\partial^{2l}}{\partial \left(  \bar{\beta}\right)  ^{l}\partial \left(
\bar{\beta}^{\ast}\right)  ^{l}}H_{m}(\bar{\beta})H_{m}(\bar{\beta}^{\ast
}).\label{3.8}%
\end{align}
Noticing the recurrence relations of $H_{m}(x)$,%
\begin{equation}
\frac{\mathtt{d}}{\mathtt{d}x^{l}}H_{m}(x)=\frac{2^{l}m!}{\left(  m-l\right)
!}H_{m-l}(x),\label{3.7}%
\end{equation}
then the Wigner function of $\left \vert r,m\right \rangle $ is given by
\begin{align}
W\left(  q,p\right)   &  =\frac{N_{m}^{2}e^{-2\left \vert \beta \right \vert
^{2}}}{2^{2m}\pi}\sinh^{m}2r\sum_{l=0}^{m}\frac{\left(  m!\right)  ^{2}\left(
-2\coth r\right)  ^{l}}{l!\left[  \left(  m-l\right)  !\right]  ^{2}%
}\left \vert H_{m-l}(\bar{\beta})\right \vert ^{2}\nonumber \\
&  =\frac{1}{\pi}\frac{e^{-2\left \vert \beta \right \vert ^{2}}\sinh^{m}r}%
{2^{m}P_{m}\left(  \cosh r\right)  }\sum_{l=0}^{m}\frac{m!\left(  -2\coth
r\right)  ^{l}}{l!\left[  \left(  m-l\right)  !\right]  ^{2}}\left \vert
H_{m-l}(\bar{\beta})\right \vert ^{2},\label{3.9}%
\end{align}
where $\beta$ and $\bar{\beta}$ are shown in Eq.(\ref{3.5b}). Eq.(\ref{3.9})
seems a new result (not reported in the literature before), related to
single-variable Hermite polynomials. In particular, when the photon-added
number $m=0,1$ and noticing that $P_{0}\left(  \cosh r\right)  =1$ and
$P_{1}\left(  \cosh r\right)  =\cosh r,$ Eq.(\ref{3.9}) reduce to,
respectively,
\begin{align}
W_{m=0}\left(  q,p\right)   &  =\frac{1}{\pi}e^{-(q^{2}e^{-2r}+p^{2}e^{2r}%
)},\label{3.10}\\
W_{m=1}\left(  q,p\right)   &  =\frac{1}{\pi}\left \{  2(q^{2}e^{-2r}%
+p^{2}e^{2r})-1\right \}  e^{-(q^{2}e^{-2r}+p^{2}e^{2r})}.\label{3.11}%
\end{align}
Eq.(\ref{3.10}) is just the WF of squeezed vacuum state, a Gaussian in phase
space; while Eq.(\ref{3.11}) corresponds to a non-Gaussian WF in phase space
due to the presence of non-Gaussian item $2(q^{2}e^{-2r}+p^{2}e^{2r})-1$. It
is clear that the function $W_{m=1}\left(  q,p\right)  $ becomes negative in
phase space, when $2(q^{2}e^{-2r}+p^{2}e^{2r})<1$.

Using Eq.(\ref{3.9}) we show the plots of WF in the phase space in Figs.2 for
different squeezing parameters $r$ and photon-added numbers $m$. One can see
clearly that there is some negative region of the WF in the phase space which
implies the nonclassicality of this state. In addition, the squeezing effect
in one of the quadratures is clear in the plots (see Figs.2a and Figs.2b),
which is another evidence of the nonclassicality of this state. The WF has its
minimum value for $m=1,3$ at the center of phase space $\left(  q=p=0\right)
$ (see Fig.2(a) and (d)). The case is not true for $m=2$ (see Fig2.c).

\begin{figure}[ptb]
\label{Fig1} \centering
\includegraphics[width=14cm]{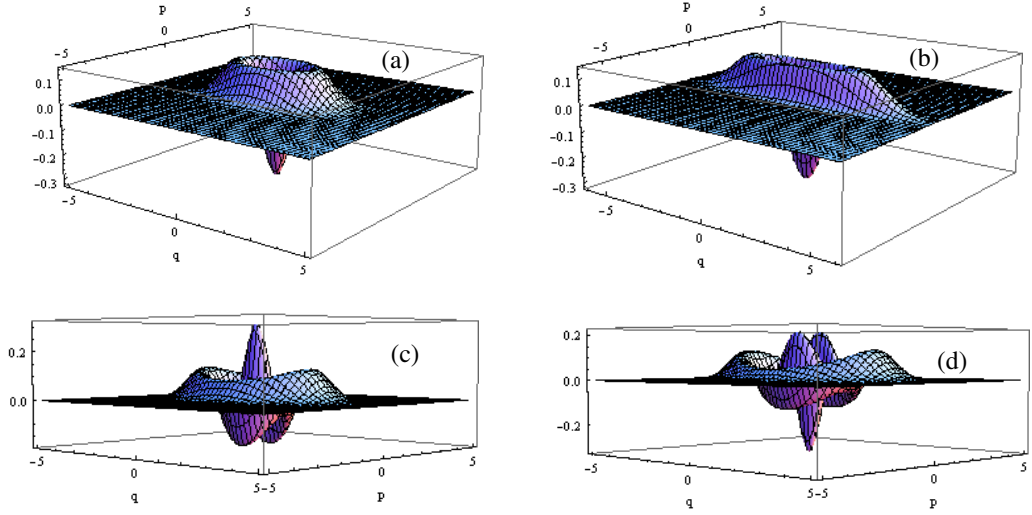}\caption{{\protect \small (Color online)
The Wigner functions of PASV for (a) }$m=1,r=0.3,${\protect \small (b)
}$m=1,r=0.8,${\protect \small (c) }$m=2,r=0.3$ {\protect \small and (d)
}$m=3,r=0.3.$}%
\end{figure}

\section{Decoherence of PASV in thermal environment}

\subsection{Model of Decoherence}

In this section, we consider how this single-mode state (\ref{2.1}) evolves at
the presence of thermal environment. In thermal channel, the evolution of the
density matrix for the $m$-PASV can be described by \cite{29}%
\begin{equation}
\frac{d\rho}{dt}=\kappa \left(  \bar{n}+1\right)  \left(  2a\rho a^{\dagger
}-a^{\dagger}a\rho-\rho a^{\dagger}a\right)  +\kappa \bar{n}\left(
2a^{\dagger}\rho a-aa^{\dagger}\rho-\rho aa^{\dagger}\right)  , \label{4.1}%
\end{equation}
where $\kappa$ represents the dissipative coefficient and $\bar{n}$ denotes
the average thermal photon number of the environment. When $\bar{n}=0,$
Eq.(\ref{4.1}) reduces to the master equation describing the photon-loss
channel. The corresponding time evolution density operator can be obtained as
\begin{equation}
\rho_{r,m}\left(  t\right)  =e^{\kappa t+\Gamma_{0}}\sum_{k,l=0}^{\infty
}M_{k,l}\rho_{0}M_{k,l}^{\dagger}, \label{4.2}%
\end{equation}
where $\rho_{0}=\left \vert r,m\right \rangle \left \langle r,m\right \vert $ is
the initial density matrix; $M_{k,l}$ and $M_{k,l}^{\dag}$ are Hermite
conjugated operators (Kraus operator) with each other,%
\begin{equation}
M_{k,l}=e^{\left(  \kappa t+\Gamma_{0}\right)  /2}\sqrt{\frac{\Gamma_{-}%
^{k}\Gamma_{+}^{l}e^{-2l\Gamma_{0}}}{k!l!}}e^{\Gamma_{0}a^{\dagger}%
a}a^{\dagger l}a^{k}, \label{4.3}%
\end{equation}
and $\Gamma_{+},\Gamma_{-}$ and $\Gamma_{0}$ are determined by
\begin{equation}
T=1-e^{-2\kappa t},\text{ }\Gamma_{+}=\frac{\bar{n}T}{\bar{n}T+1},\text{
}\Gamma_{-}=\frac{\left(  \bar{n}+1\right)  T}{\bar{n}T+1},\text{ }\Gamma
_{0}=\ln \frac{e^{-\kappa t}}{\bar{n}T+1}. \label{4.4}%
\end{equation}
It is not difficult to prove the $M_{k,l}$ obeys the normalization condition
$\sum_{k,l=0}^{\infty}M_{k,l}^{\dag}M_{k,l}=1$ by using the technique of
integration within an ordered products of operators \cite{30,31}.

\subsection{Wigner function of the PASV in a thermal channel}

The evolution formula of WF of the PASV can be derived as follows \cite{31a}
\begin{equation}
W\left(  \zeta,\zeta^{\ast},t\right)  =\frac{2}{\left(  2\bar{n}+1\right)
T}\int \frac{d^{2}\alpha}{\pi}W\left(  \alpha,\alpha^{\ast},0\right)
e^{-2\frac{\allowbreak \left \vert \zeta-\alpha e^{-\kappa t}\right \vert ^{2}%
}{\left(  2\allowbreak \bar{n}+1\right)  T}},\label{4.5}%
\end{equation}
where $W\left(  \alpha,\alpha^{\ast},0\right)  $ is the WF of the initial
state. Eq.(\ref{4.5}) is just the evolution formula of WF in thermal channel.
Thus the WF at any time can be obtained by performing the integration when the
initial WF is known.

In a similar way to deriving Eq.(\ref{3.9}), substituting Eq.(\ref{3.9}) into
Eq.(\ref{4.5}) and using the generating function of single-variable Hermite
polynomials (\ref{2.5}) and Eq.(\ref{3.6}), we finally obtain (see appendix B)%
\begin{align}
W\left(  \zeta,\zeta^{\ast},t\right)   &  =\frac{2\sinh^{m}r}{\pi \sqrt
{C}\left(  2\bar{n}+1\right)  T}\frac{m!e^{-\frac{2\left \vert \zeta \right \vert
^{2}}{\left(  2\allowbreak \bar{n}+1\right)  T}}}{2^{m}P_{m}\left(  \cosh
r\right)  }e^{\frac{A}{C}\left \vert B\right \vert ^{2}+\frac{\allowbreak
\sinh2r}{C}\left(  B^{\ast}{}^{2}+B{}^{2}\right)  }\nonumber \\
&  \times \sum_{l=0}^{m}\sum_{k=0}^{m-l}\frac{\left(  -2\coth r\right)
^{l}G^{m-l-k}F^{k}}{l!k!\left[  \left(  m-l-k\right)  !\right]  ^{2}%
}\left \vert H_{m-l-k}(E/\sqrt{G})\right \vert ^{2}.\label{4.6}%
\end{align}
where we have set
\begin{align}
A &  =\frac{2e^{-2\kappa t}}{\left(  2\allowbreak \bar{n}+1\right)  T}%
+2\cosh2r,\text{ }\label{4.7}\\
B &  =\frac{2e^{-\kappa t}\zeta}{\left(  2\allowbreak \bar{n}+1\right)
T},\text{ }C=A^{2}-4\sinh^{2}2r,\text{ }\label{4.8}\\
D &  =\sqrt{2\coth r}i\left(  B^{\ast}\sinh r-B\cosh r\right)  ,\label{4.9}\\
E &  =\frac{1}{C}\left(  AD-2D^{\ast}\sinh2r\right)  ,\label{4.10}\\
F &  =\frac{8}{C}\left(  A\cosh2r\coth r-4\cosh^{2}r\sinh2r\right)
,\label{4.11}\\
G &  =1-\frac{16}{C}\frac{e^{-2\kappa t}}{\left(  2\allowbreak \bar
{n}+1\right)  T}\cosh^{2}r.\label{4.12}%
\end{align}
Eq.(\ref{4.6}) is just the analytical expression of WF for the PASV in thermal
channel. It is obvious that the WF loss its Gaussian property due to the
presence of single-variable Hermite polynomials. In particular, at the initial
time ($t=0$)$,$ noting $\sqrt{C}\left(  2\bar{n}+1\right)  T\rightarrow2,$
$G\rightarrow1,$ $F/G\rightarrow0,$ and $E/\sqrt{G}\rightarrow \bar{\beta
}=-i\sqrt{2\coth r}\left(  \zeta \cosh r-\zeta^{\ast}\sinh r\right)  ,$ as well
as $\frac{A}{C}\left \vert B\right \vert ^{2}-\frac{2\left \vert \zeta \right \vert
^{2}}{\left(  2\allowbreak \bar{n}+1\right)  T}\rightarrow-2\left \vert
\zeta \right \vert ^{2}\cosh2r,\frac{\allowbreak \sinh2r}{C}\left(  B^{\ast}%
{}^{2}+B{}^{2}\right)  \rightarrow \left(  \zeta^{2}+\zeta^{\ast2}\right)
\sinh2r$, Eq.(\ref{4.6}) just does reduce to Eq.(\ref{3.9}), i.e., the WF of
the PASV. On the other hand, when $\kappa t\rightarrow \infty,$ noticing that
$T\rightarrow1,B\rightarrow0,C\rightarrow4,D\rightarrow0,$ $E/\sqrt
{G}\rightarrow0,G\rightarrow1,$ $F\rightarrow \allowbreak4\coth r,$as well as
$H_{m}\left(  0\right)  =\left(  -1\right)  ^{j}\frac{m!}{j!}\delta_{m,2j},$
then Eq.(\ref{4.6}) becomes $\allowbreak \frac{1}{\pi \left(  2\bar{n}+1\right)
}e^{-\frac{2\left \vert \zeta \right \vert ^{2}}{2\allowbreak \bar{n}+1}}$, which
is independent of photon-addition number $m$ and corresponds to the WF of
thermal state with mean thermal photon number $\bar{n}$. This indicates that
the system state reduces to a thermal state after an enough long time
interaction with the environment.

In addition, for the case of $m=0,$ single-mode squeezed vacuum,
Eq.(\ref{4.6}) just becomes ($H_{0}(x)=1$)%
\begin{equation}
W_{m=0}\left(  \zeta,\zeta^{\ast},t\right)  =\mathfrak{N}e^{-\mathfrak{D}%
\left \vert \zeta \right \vert ^{2}+\mathfrak{E(}\zeta^{\ast}{}^{2}+\zeta{}^{2}%
)},\label{4.14}%
\end{equation}
where $\mathfrak{N}=\frac{2}{\pi \sqrt{C}\left(  2\bar{n}+1\right)  T}$ is the
normalization factor, $\mathfrak{D}=\frac{2}{\left(  2\allowbreak \bar
{n}+1\right)  T}-\frac{A\mathfrak{E}}{\sinh2r},\mathfrak{E}\mathfrak{=}%
\frac{4e^{-2\kappa t}\sinh2r}{\left[  \left(  2\allowbreak \bar{n}+1\right)
T\right]  ^{2}C},$ Eq.(\ref{4.14}) denotes a Gaussian distribution function---
the WF of single-mode squeezed vacuum in the thermal channel; while for $m=1$,
single--photon added case, its WF in the thermal channel is given by
($H_{1}(x)=2x$)
\begin{equation}
W_{m=1}\left(  \zeta,\zeta^{\ast},t\right)  =\frac{4\left \vert E\right \vert
^{2}+F-2\coth r}{\pi \sqrt{C}\left(  2\bar{n}+1\right)  T\coth r}e^{\frac{A}%
{C}\left \vert B\right \vert ^{2}-\frac{2\left \vert \zeta \right \vert ^{2}%
}{\left(  2\allowbreak \bar{n}+1\right)  T}+\frac{\allowbreak \sinh2r}{C}\left(
B^{\ast}{}^{2}+B{}^{2}\right)  }.\label{4.15}%
\end{equation}

In Fig.3, the WFs of the PASV with $m=1$ are depicted in phase space with
$r=0.3$ and $\bar{n}=1$ for several different $\kappa t.$ It is easy to see
that the negative region of WF gradually diminishes as the time $\kappa t$
increases. Actually, from Eq.(\ref{4.8}) one can see that $C>0$, so at the
center of the phase space ($\alpha=\alpha^{\ast}=0$), when $F<2\coth r$
leading to the following condition:%
\begin{equation}
\kappa t<\kappa t_{c}\equiv \frac{1}{2}\ln \frac{2\bar{n}+2}{2\bar{n}%
+1},\label{4.16}%
\end{equation}
which is independent of the squeezing parameter $r$, there always exist
negative region for WF in phase space and the WF of PASV is always positive in
the whole phase space when $\kappa t\ $exceeds the threshold value $\kappa
t_{c}$.

In Figs. 4 and 5, we plot the variation of WF in phase space for different
$\bar{n}$ and $r,$ respectively. It is found that the partial negativity of WF
decreases gradually as $\bar{n}$ (or $r$) increases for a given time. The
squeezing effect in one of the quadrature is shown in Fig.5. In addition, for
the case of large squeezing value $r$, the single-photon added squeezed state
becomes similar to a Schodinger cat state (see Fig.6). The WF becomes Gaussian
with the time evolution. In principle, by using the explicit expression of WF
in Eq.(\ref{4.6}), we can draw the WF distribution for any photon-added case
in phase space. For instance, for $m=2,$ there are two negative regions of the
WF, which differs from the case of single PASV (see Fig.7). The absolute value
of the negative minimum of the WF decreases as $\kappa t$ increases, which
leads to the complete absence of partial negative region.

\section{Conclusions}

The nonclassical properties and decoherence of single-mode PASV in a thermal
environment have been investigated. A compact expression for the normalization
factor of PASV is derived by converting the PASV to a squeezed Hermite
polynomial excitation state. It is shown that the normalization factor is just
an $m$-order Legendre polynomial of the squeezing parameter $r$. We also
derived the explicit analytical expression of WF for any photon-added number
$m$ and found the negative region of WF in phase space. We also show that
there is an upper bound value of $r$ for this state to exhibit sub-Poissonian
statistics which increases as $m$ increases. Then we considered the effects of
decoherence to the nonclassicality of PASV when interacting with thermal
environment. For arbitrary number PASV, we derived the explicit analytical
expression of time evolution of WF and presented the loss of nonclassicality
in reference of the negativity of WF. The threshold value of the decay time
corresponding to the transition of the WF from partial negative to completely
positive definite is obtained. We find that the WF has always negative value
for all parameters $r$ if the decay time $\kappa t<\frac{1}{2}\ln \frac
{2\bar{n}+2}{2\bar{n}+1}$ for single PASV.

\bigskip

\textbf{ACKNOWLEDGEMENTS:} Work supported by the National Natural Science
Foundation of China (Grant Nos 10775097 and 10874174) and the Research
Foundation of the Education Department of Jiangxi Province.
\begin{figure}[ptb]
\label{Fig2}
\centering \includegraphics[width=14cm]{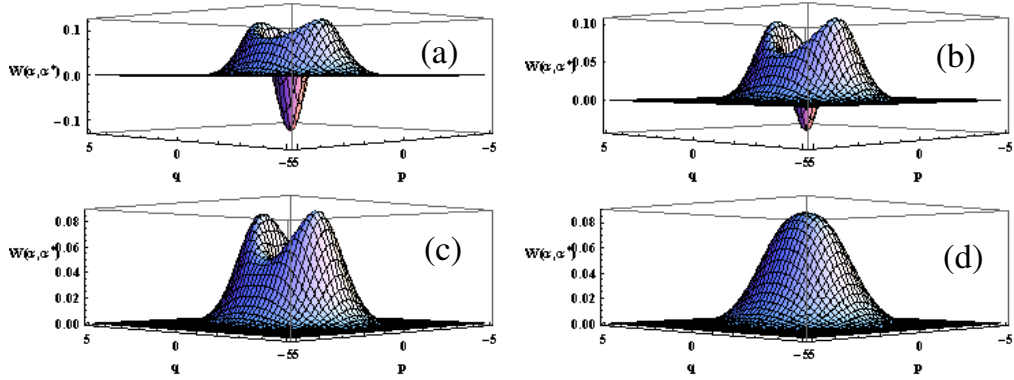}\caption{{\protect \small (Color
online) The Wigner functions of single-photon-added squeezed vacuum states in
phase space for }$r=0.3,\bar{n}=1${\protect \small \ at }$(a)$%
{\protect \small \ }$\kappa t=0.05,(b)${\protect \small \ }$\kappa
t=0.1,(c)${\protect \small \ }$\kappa t=0.2\ ${\protect \small and }%
$(d)${\protect \small \ }$\kappa t=0.5.$}%
\end{figure}\begin{figure}[ptb]
\label{Fig3}
\centering \includegraphics[width=14cm]{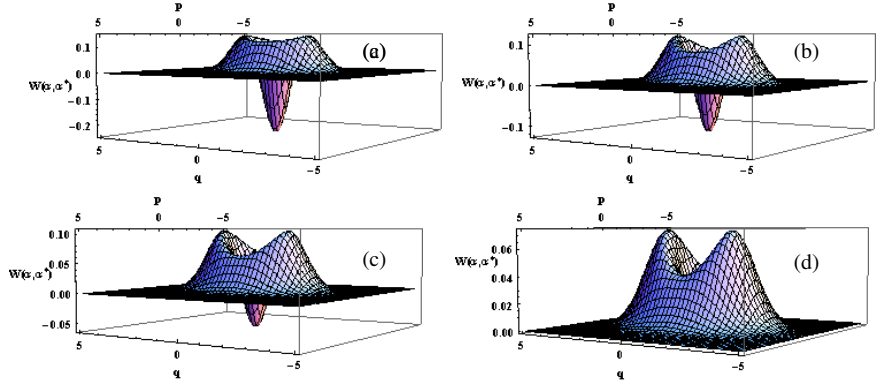}\caption{{\protect \small (Color
online) The Wigner functions of single-photon-added squeezed vacuum states in
phase space for }$r=0.3${\protect \small \ and }$\kappa t=0.05$%
{\protect \small \ with }$(a)${\protect \small \ }$\bar{n}=0,(b)$%
{\protect \small \ }$\bar{n}=1,(c)${\protect \small \ }$\bar{n}=2,(d)$%
{\protect \small \ }$\bar{n}=5.$}%
\end{figure}\begin{figure}[ptb]
\label{Fig4}
\centering \includegraphics[width=14cm]{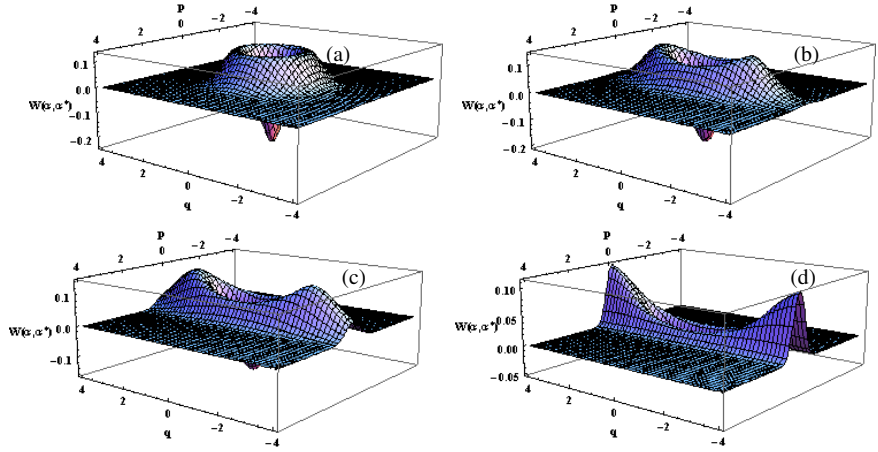}\caption{{\protect \small (Color
online) The Wigner functions of single-photon-subtracted squeezed vacuum
states in phase space for }$\bar{n}=0.1${\protect \small \ and }$\kappa
t=0.05${\protect \small \ with }$(a)${\protect \small \ }$r=0.03,(b)$%
{\protect \small \ }$r=0.5,(c)${\protect \small \ }$r=0.8,(d)$%
{\protect \small \ }$r=1.5.??$}%
\end{figure}\begin{figure}[ptb]
\label{Fig5}
\centering \includegraphics[width=14cm]{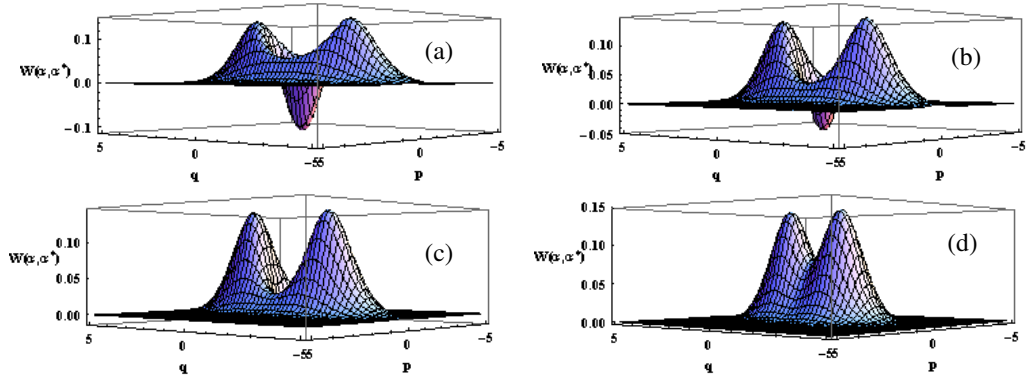}\caption{{\protect \small (Color
online) Wigner function of single photon-added squeezed vacuum states in phase
space for }$r=0.8,\bar{n}=0${\protect \small : }$(a)${\protect \small \ }$\kappa
t=0.1,(b)${\protect \small \ }$\kappa t=0.2,(c)${\protect \small \ }$\kappa
t=0.3,\ ${\protect \small and }$(d)${\protect \small \ }$\kappa t=0.7.$}%
\end{figure}\begin{figure}[ptb]
\label{Fig6}
\centering \includegraphics[width=14cm]{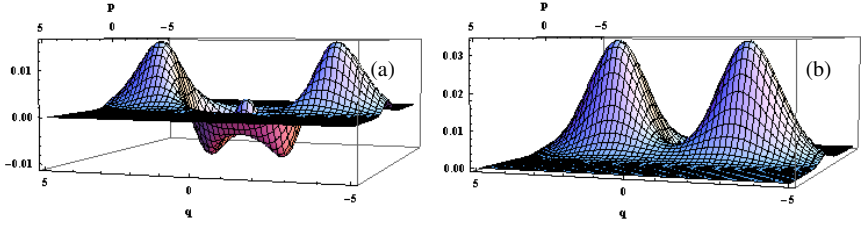}\caption{{\protect \small (Color
online) The Wigner functions of photon-added squeezed vacuum states in phase
space for }$r=0.7,\bar{n}=1,m=2${\protect \small : }$(a)${\protect \small \ }%
$\kappa t=0.1,(b)${\protect \small \ }$\kappa t=0.2.$}%
\end{figure}

\newpage

\textbf{Appendix A: Derivation of Eq.(\ref{2.7})}

Recalling that the newly found expression of Legendre polynomial \cite{20,21}
\begin{equation}
x^{m}\sum_{{l}=0}^{\left[  m/2\right]  }\frac{m!}{2^{2{l}}\left(  {l}!\right)
^{2}\left(  m-2{l}\right)  !}\left(  1-\frac{1}{x^{2}}\right)  ^{{l}}%
=P_{m}\left(  x\right)  , \tag{A1}%
\end{equation}
which is equivalent to the well-known Legendre polynomial's expression
\cite{32}
\begin{equation}
P_{m}\left(  x\right)  =\sum_{l=0}^{[m/2]}\left(  -1\right)  ^{l}\frac{\left(
2m-2l\right)  !}{2^{m}l!\left(  m-l\right)  !\left(  m-2l\right)  !}x^{m-2l},
\tag{A2}%
\end{equation}
though it is different in form from Eq.(A1), they actually are equal. For
instance, what we list in the following equations (the left) from Eq.(A1) are
equal to what we list from (A2) (the right)%
\begin{align}
m  &  =0,\text{ }P_{0}\left(  x\right)  =1;\text{ \ }\nonumber \\
\text{\ }m  &  =1,\text{ }P_{1}\left(  x\right)  =x;\nonumber \\
m  &  =2,\text{ }P_{2}\left(  x\right)  =x^{2}\left[  1+\frac{1}{2}\left(
1-\frac{1}{x^{2}}\right)  \right]  =\frac{3}{2}x^{2}-\frac{1}{2};\nonumber \\
m  &  =3,\text{ }P_{3}\left(  x\right)  =x^{3}\left[  1+\frac{3}{2}\left(
1-\frac{1}{x^{2}}\right)  \right]  =\frac{5}{2}x^{3}-\frac{3}{2}x; \tag{A3}%
\end{align}
and%
\begin{align}
m  &  =4,\text{ }P_{4}\left(  x\right)  =x^{4}\left[  1+3\left(  1-\frac
{1}{x^{2}}\right)  +\frac{3}{8}\left(  1-\frac{1}{x^{2}}\right)  ^{{2}%
}\right]  =\frac{1}{8}\left(  35x^{4}-30x^{2}+3\right)  ;\nonumber \\
m  &  =5,\text{ }P_{5}\left(  x\right)  =x^{5}\left[  1+5\left(  1-\frac
{1}{x^{2}}\right)  +\allowbreak \frac{15}{8}\left(  1-\frac{1}{x^{2}}\right)
^{{2}}\right]  =\frac{1}{8}\left(  63x^{5}-70x^{3}+15x\right)  ; \tag{A4}%
\end{align}
We emphasize that the new form in Eq.(A1) cannot be directly obtained by some
series summation rearrangement technique from the orginal definition (A2).

On the other hand, noting the following relation%
\begin{align}
&  \frac{\partial^{2m}}{\partial t^{m}\partial \tau^{m}}\left.  \exp \left(
-t^{2}-\tau^{2}+2x\tau t\right)  \right \vert _{t,\tau=0}\nonumber \\
&  =\sum_{n,l,k=0}^{\infty}\frac{\left(  -\right)  ^{n+l}}{n!l!k!}\left(
2x\right)  ^{k}\left.  \frac{\partial^{2m}}{\partial t^{m}\partial \tau^{m}%
}\tau^{2n+k}t^{2l+k}\right \vert _{t,\tau=0}\nonumber \\
&  =2^{m}m!\sum_{n=0}^{\left[  m/2\right]  }\frac{m!}{2^{2n}\left(  n!\right)
^{2}\left(  m-2n\right)  !}x^{m-2n}, \tag{A5}%
\end{align}
and then comparing Eq.(A5) with Eq.(A1) we can obtain Eq.(\ref{2.7}).

\textbf{APPENDIX B: Derivation of Wigner function Eq.(\ref{4.6}) of PASV}

In this appendix, we present the details for deriving the Wigner function
Eq.(\ref{4.6}). Substituting Eq.(\ref{3.9}) into Eq.(\ref{4.5}) and noticing
Eq.(\ref{3.5b}) as well as the generating function of single-variable Hermite
polynomials (\ref{2.5}), we have%
\begin{align}
W\left(  \zeta,\zeta^{\ast},t\right)   &  =\bar{N}\sum_{l=0}^{m}\frac{\left(
-2\coth r\right)  ^{l}}{l!\left[  \left(  m-l\right)  !\right]  ^{2}}\int
\frac{d^{2}\alpha}{\pi}e^{-2\frac{\allowbreak \left \vert \zeta-\alpha
e^{-\kappa t}\right \vert ^{2}}{\left(  2\allowbreak \bar{n}+1\right)
T}-2\left \vert \alpha \cosh r-\allowbreak \alpha^{\ast}\sinh r\right \vert ^{2}%
}\nonumber \\
&  \times \left \vert H_{m-l}(-i\left(  \alpha \cosh r-\allowbreak \alpha^{\ast
}\sinh r\right)  \sqrt{2\coth r})\right \vert ^{2}\nonumber \\
&  =\bar{N}e^{-\frac{2\left \vert \zeta \right \vert ^{2}}{\left(  2\allowbreak
\bar{n}+1\right)  T}}\sum_{l=0}^{m}\frac{\left(  -2\coth r\right)  ^{l}%
}{l!\left[  \left(  m-l\right)  !\right]  ^{2}}\frac{\partial^{2m-2l}%
}{\partial \upsilon^{m-l}\partial \tau^{m-l}}e^{-\upsilon^{2}-\tau^{2}%
}\nonumber \\
&  \times \int \frac{d^{2}\alpha}{\pi}\left.  \exp \left \{  -A\left \vert
\alpha \right \vert ^{2}+B_{1}\alpha+B_{2}\alpha^{\ast}+\left(  \allowbreak
\alpha^{2}+\alpha^{\ast}{}^{2}\right)  \sinh2r\right \}  \right \vert
_{\tau=\upsilon=0}, \tag{B1}%
\end{align}
where $A$ and $B$ are given by Eqs.(\ref{4.9})-(\ref{4.10}),%
\begin{equation}
\bar{N}=\frac{2\sinh^{m}r}{\pi \left(  2\bar{n}+1\right)  T}\frac{m!}%
{2^{m}P_{m}\left(  \cosh r\right)  }, \tag{B2}%
\end{equation}
and%
\begin{align}
B_{1}  &  =B^{\ast}\allowbreak-i2\left(  \upsilon \cosh r+\tau \sinh r\right)
\sqrt{2\coth r},\tag{B3}\\
B_{2}  &  =B+i2\allowbreak \left(  \upsilon \sinh r+\tau \cosh r\right)
\sqrt{2\coth r}. \tag{B4}%
\end{align}
Further using the following integral identity \cite{33}:%
\begin{align}
&  \int \frac{d^{2}z}{\pi}\exp \left(  \zeta \left \vert z\right \vert ^{2}+\xi
z+\eta z^{\ast}+fz^{2}+gz^{\ast2}\right) \nonumber \\
&  =\frac{1}{\sqrt{\zeta^{2}-4fg}}\exp \left[  \frac{-\zeta \xi \eta+\xi
^{2}g+\eta^{2}f}{\zeta^{2}-4fg}\right]  , \tag{B5}%
\end{align}
whose convergent condition is Re$\left(  \zeta \pm f\pm g\right)  <0,\ $and
Re$\left(  \frac{\zeta^{2}-4fg}{\zeta \pm f\pm g}\right)  <0,$ we can put
Eq.(B1) into the following form:%
\begin{align}
W\left(  \zeta,\zeta^{\ast},t\right)   &  =\frac{\bar{N}}{\sqrt{C}}%
e^{-\frac{2\left \vert \zeta \right \vert ^{2}}{\left(  2\allowbreak \bar
{n}+1\right)  T}}\sum_{l=0}^{m}\frac{\left(  -2\coth r\right)  ^{l}}{l!\left[
\left(  m-l\right)  !\right]  ^{2}}\frac{\partial^{2m-2l}}{\partial
\upsilon^{m-l}\partial \tau^{m-l}}\nonumber \\
&  \times \exp \left \{  -\upsilon^{2}-\tau^{2}+\frac{A}{C}B_{1}B_{2}%
+\frac{\allowbreak1}{C}\left(  B_{1}^{2}+B_{2}^{2}\right)  \sinh2r\right \}
_{\tau=\upsilon=0}, \tag{B6}%
\end{align}
where $C$ and $D$ are given by (\ref{4.9})-(\ref{4.10}), and
\begin{align}
B_{1}B_{2}  &  =B^{\ast}B+2\left(  \allowbreak \upsilon D+\tau D^{\ast}\right)
+8\left(  \tau^{2}+\upsilon^{2}\right)  \cosh^{2}r+8\tau \upsilon \cosh2r\coth
r,\tag{B7}\\
B_{1}^{2}+B_{2}^{2}  &  =B^{\ast}{}^{2}+B{}^{2}-32\tau \upsilon \cosh
^{2}r-8\left(  \upsilon^{2}+\tau^{2}\right)  \cosh2r\coth r-\allowbreak4\tau
D-4\upsilon D^{\ast}, \tag{B8}%
\end{align}
then substituting Eqs.(B7)-(B8) into Eq.(B6) yields%
\begin{align}
W\left(  \zeta,\zeta^{\ast},t\right)   &  =\frac{\bar{N}}{\sqrt{C}}%
e^{-\frac{2\left \vert \zeta \right \vert ^{2}}{\left(  2\allowbreak \bar
{n}+1\right)  T}}e^{\frac{A}{C}\left \vert B\right \vert ^{2}+\frac
{\allowbreak \sinh2r}{C}\left(  B^{\ast}{}^{2}+B{}^{2}\right)  }\sum_{l=0}%
^{m}\frac{\left(  -2\coth r\right)  ^{l}}{l!\left[  \left(  m-l\right)
!\right]  ^{2}}\nonumber \\
&  \times \frac{\partial^{2m-2l}}{\partial \upsilon^{m-l}\partial \tau^{m-l}}%
\exp \left \{  -G\left(  \upsilon^{2}+\tau^{2}\right)  +\allowbreak2\upsilon
E+2\tau E^{\ast}+\tau \upsilon F\right \}  _{\tau=\upsilon=0}, \tag{B9}%
\end{align}
where $E,F,$ and $G$ are given by Eqs.(\ref{4.10})-(\ref{4.12}). In order to
further simplify Eq.(B9), expanding the exponential item $e^{\tau \upsilon F}$
as sum series and using Eqs.(\ref{2.5}) and the formula
\begin{equation}
\left.  \frac{\partial^{m}}{\partial \upsilon^{m}}e^{-G\upsilon^{2}%
+\allowbreak2\upsilon E}\right \vert _{\upsilon=0}=G^{m/2}H_{m}\left(
E/\sqrt{G}\right)  , \tag{B10}%
\end{equation}
we see%
\begin{align}
W\left(  \zeta,\zeta^{\ast},t\right)   &  =\frac{\bar{N}}{\sqrt{C}}%
e^{-\frac{2\left \vert \zeta \right \vert ^{2}}{\left(  2\allowbreak \bar
{n}+1\right)  T}+\frac{A}{C}\left \vert B\right \vert ^{2}+\frac{\allowbreak
\sinh2r}{C}\left(  B^{\ast}{}^{2}+B{}^{2}\right)  }\sum_{l=0}^{m}\frac{\left(
-2\coth r\right)  ^{l}}{l!\left[  \left(  m-l\right)  !\right]  ^{2}%
}\nonumber \\
&  \times \sum_{k=0}^{\infty}\frac{F^{k}}{k!}\frac{\partial^{2k}}%
{\partial \left(  2E\right)  ^{k}\partial \left(  2E^{\ast}\right)  ^{k}}%
\frac{\partial^{2m-2l}}{\partial \upsilon^{m-l}\partial \tau^{m-l}}\left.
e^{-G\left(  \upsilon^{2}+\tau^{2}\right)  +\allowbreak2\upsilon E+2\tau
E^{\ast}}\right \vert _{\tau=\upsilon=0}\nonumber \\
&  =\frac{\bar{N}}{\sqrt{C}}e^{-\frac{2\left \vert \zeta \right \vert ^{2}%
}{\left(  2\allowbreak \bar{n}+1\right)  T}+\frac{A}{C}\left \vert B\right \vert
^{2}+\frac{\allowbreak \sinh2r}{C}\left(  B^{\ast}{}^{2}+B{}^{2}\right)  }%
\sum_{l=0}^{m}\frac{\left(  -2\coth r\right)  ^{l}G^{m-l}}{l!\left[  \left(
m-l\right)  !\right]  ^{2}}\nonumber \\
&  \times \sum_{k=0}^{\infty}\frac{F^{k}}{k!}\frac{\partial^{2k}}%
{\partial \left(  2E\right)  ^{k}\partial \left(  2E^{\ast}\right)  ^{k}%
}\left \vert H_{m-l}\left(  E/\sqrt{G}\right)  \right \vert ^{2}. \tag{B11}%
\end{align}
Then using Eq.(\ref{3.7}) yields Eq.(\ref{4.6}).


\begin{thebibliography}{99}                                                                                               %


\bibitem {1}D. Bouwmeester, A. Ekert and A. Zeilinger, \textit{The Physics of
Quantum Information} (Springer-Verlag, Berlin, 2000).

\bibitem {2}R. Short and L. Mandel, Phys. Rev. Lett. \textbf{51}, 384 (1983).

\bibitem {3}V. V. Dodonov, J. Opt. B: Quantum Semiclassical Opt. \textbf{4},
R1 (2002).

\bibitem {4}M. Hillery, R. F. O'Connell, M. O. Scully, and E. P. Wigner, Phys.
Rep. \textbf{106}, 121 (1984).

\bibitem {5}J. Wenger, R. Tualle-Brouri, and P. Grangier, Phys. Rev. Lett. 92,
153601 (2004).

\bibitem {6}A. Zavatta, S. Viciani, and M. Bellini, Science, \textbf{306}, 660 (2004).

\bibitem {7}A. Zavatta, S. Viciani, and M. Bellini, Phys. Rev. A \textbf{72},
023820 (2005).

\bibitem {8}V. Parigi, A. Zavatta, M. S. Kim, and M. Bellini, Science,
\textbf{317}, 1890 (2007).

\bibitem {8a}R. W. Boyd, K. W. Chan, and M. N. O'Sullivan, Science,
\textbf{317}, 1874 (2007).

\bibitem {9}A. Zavatta, S. Viciani, and M. Bellini, Phys. Rev. A \textbf{75},
052106 (2007).

\bibitem {10}J. S. Neergaard-Nielsen, B. Melholt Nielsen, C. Hettich, K.
M\o lmer, and E. S. Polzik,\ Phys. Rev. Lett. 97, 083604 (2006).

\bibitem {11}A. Ourjoumtsev, R. Tualle-Brouri, J. Laurat, Ph.
Grangier,\ Science \textbf{312}, 83 (2006).

\bibitem {12}K. Wakui, H. Takahashi, A. Furusawa, and M. Sasaki,\ Opt. Express
\textbf{15}, 3568 (2007).

\bibitem {13}M. Dakna, T. Anhut, T. Opatrny, L. Knoll, and D.-G.
Welsch,\ Phys. Rev. A \textbf{55}, 3184 (1997).

\bibitem {14}S. Glancy and H. M. de Vasconcelos, J. Opt. Soc. Am. B
\textbf{25}, 712 (2008).

\bibitem {15}A. Biswas and G. S. Agarwal,\ Phys. Rev. A \textbf{75}, 032104 (2007).

\bibitem {16}H. Jeong, A. P. Lund, and T. C. Ralph, Phys. Rev. A \textbf{72},
013801 (2005).

\bibitem {16a}B. T. H. Vracoe, S. Brattke, M. Weidinger, H. Walther, Nature
\textbf{403}, 743 (2000);

S. Brattke, B.T.H. Vracoe, H. Walther, Phys. Rev. Lett. \textbf{86}, 3534 (2001).

\bibitem {16b}Z. M. Zhang, Chin. Phys. Lett. 20 (2003) 227;

Z. M. Zhang, Chin. Phys. Lett. 21 (2004) 5.

\bibitem {16c}Z. X. Zhang, H. Y. Fan, Phys. Lett. A 174 (1993) 206.

\bibitem {17}D. F. Walls andG J Milburn, \textit{Quantum Optics}
(Springer-Verlag, Berlin, 1994).

\bibitem {18}M. O. Scully and M. S. Zubairy, \textit{Quantum Optics}
(Cambridge: Cambidge University Press, 1997).

\bibitem {19}H.Y. Fan and Vander J. Linde, J. Phys. A \textbf{24}, 2529 (1989).

\bibitem {20}L. Y. Hu and H. Y. Fan, J. Opt. Soc. Am. B, \textbf{25}, 1955 (2008).

\bibitem {21}H. Y. Fan, X. G. Meng and J. S. Wang, Commun. Theor. Phys.
\textbf{46,} 845 (2006).

\bibitem {23}M. S. Kim, J. Phys. B: At. Mol. Opt. Phys. \textbf{41}, 133001 (2008).

\bibitem {24}P. Marek, H. Jeong, and M. S. Kim, Phys. Rev. A 78, 063811 (2008).

\bibitem {25}H.-Y. Fan, H. R. Zaidi, Phys. Lett. A 124, 303 (1987).

\bibitem {26}H.-Y. Fan, Representation and Transformation Theory in Quantum
Mechanics, Shanghai Scientific \& Technical, Shanghai Press, 1997.

\bibitem {27}R. Glauber, Phys. Rev. \textbf{130,} 2529 (1963).

\bibitem {28}R. Glauber,\ Phys. Rev. \textbf{131,} 2766 (1963).

\bibitem {29}C. Gardiner and P. Zoller, \textit{Quantum Noise} (Springer,
Berlin, 2000).

\bibitem {30}H.-Y. Fan, H.-L. Lu and Y. Fan, Ann. Phys. \textbf{321,} 480 (2006).

\bibitem {31}H.-Y. Fan and L. Y. Hu, Mod. Phys. Lett. B, 22, 2435 (2008).

\bibitem {31a}L. Y. Hu and H.-Y. Fan, Opt. Commun. \textbf{282}, 4379 (2009).

\bibitem {32}I. S. Gradshteyn and L. M. Ryzhik, \textit{Tables of Integration
Series and Products} (Academic Press, New York, 1980).

\bibitem {33}R. R. Puri, \textit{Mathematical Methods of Quantum Optics}
(Springer-Verlag, Berlin, 2001), Appendix A.
\end{thebibliography}
\end{document}